\documentclass[12pt,titlepage,acknowledgements,a4paper]{article}
\usepackage{amsmath}
\usepackage{amsfonts}

\newcommand{\bp}{\bar{\psi}}
\newcommand{\e}{\epsilon}
\newcommand{\be}{\bar{\epsilon}}
\newcommand{\g}{\Gamma}

\newcommand{\nn}{\nonumber}

\newcommand{\Tr}{\textrm{Tr}}

\newcommand{\mud}{\dot{\mu}}
\newcommand{\nud}{\dot{\nu}}
\newcommand{\rhod}{\dot{\rho}}
\newcommand{\sigmad}{\dot{\sigma}}
\newcommand{\lambdad}{\dot{\lambda}}
\newcommand{\omegad}{\dot{\omega}}
\newcommand{\deltad}{\dot{\delta}}
\newcommand{\alphad}{\dot{\alpha}}
\newcommand{\betad}{\dot{\beta}}
\newcommand{\gammad}{\dot{\gamma}}

\newcommand{\mH}{\mathcal{H}}
\newcommand{\D}{\mathcal{D}}
\newcommand{\tD}{\hat{D}}
\newcommand{\tF}{\hat{F}}
 
\newcommand{\la}{\langle}
\newcommand{\ra}{\rangle}

\makeatletter
\@addtoreset{equation}{section}
\makeatother

\newcounter{multieqs}

\begin{document}

\begin{flushright}
QMUL-PH-09-19
\end{flushright}

\vspace{20pt}

\begin{center}

{\Large \bf Worldvolume Superalgebra Of BLG Theory With Nambu-Poisson Structure}\\
\vspace{33pt}

{\bf {\mbox{Andrew M. Low}}}%
\footnote{{{\tt a.m.low}{\tt @qmul.ac.uk }}}

{\em Centre for Research in String Theory \\ Department of Physics\\
Queen Mary, University of
London\\
Mile End Road, London, E1 4NS\\
United Kingdom}
\vspace{40pt}

{\bf Abstract}

\end{center}

\noindent
Recently it was proposed that the Bagger-Lambert-Gustavsson theory with Nambu-Poisson structure describes an M5-brane in a three-form flux background. In this paper we investigate the superalgebra associated with this theory. We derive the central charges corresponding to M5-brane solitons in 3-form backgrounds. We also show that double dimensional reduction of the superalgebra gives rise to the Poisson bracket terms of a non-commutative D4-brane superalgebra. We provide interpretations of the D4-brane charges in terms of spacetime intersections.
\vspace{0.5cm}

\setcounter{page}{0}
\thispagestyle{empty}
\newpage


\setcounter{footnote}{0}

\section{Introduction}

M-theory contains two basic extended objects which preserve half the supersymmetry of the vacuum; the M2-brane and the M5-brane. This is due to the fact that 11-dimensional supergravity, which is the low energy limit of M-theory, contains a single 3-form gauge field, which couples electrically to the M2-brane and magnetically to the M5-brane. The existence of these objects can also be seen at the level of the M-theory superalgebra, which is a modification of the 11-dimensional supertranslation algebra to include 2-form and 5-form central charges (For a review see \cite{Towns}). M-theory configurations which preserve less than half the supersymmetry of the vacuum may be found by looking at intersecting M-branes. The simplest case is when two M-branes intersect to yield a quarter supersymmetric solution. A remarkable feature of these configurations is that the intersection appears as a half supersymmetric soliton of the worldvolume theory of one of the constituent branes \cite{Callan,Gibb}. The spacetime interpretation can be deduced from the worldvolume scalars which determine the spacetime embedding. Furthermore it was shown in \cite{Berg} that the spacetime interpretation of a worldvolume p-brane preserving 1/2 supersymmetry is encoded in the worldvolume supersymmetry algebra. For example, in M-theory, a stack of multiple M2-branes ending on an M5-brane can be seen from the M5-brane worldvolume perspective as a self-dual string soliton \cite{Howe}, whereas the M5-M5 intersection can be described as a 3-brane vortex on the worldvolume of one of the M5-branes \cite{Howe2}. Central charges corresponding to both these configurations can be found in the worldvolume superalgebra of the M5-brane \cite{Soro, Gaunt}. For a general review of M-branes and their interactions see \cite{Berman2}.

It should also be possible to see the M5-brane as a soliton in the worldvolume theory of multiple M2-branes. Although the worldvolume theories of a single M2-brane and M5-brane are quite well understood, it is only recently that effort has been directed toward a Lagrangian description of the worldvolume theory for multiple M2-branes. This work began with the efforts of Bagger, Lambert and Gustavson \cite{BL1}-\cite{GUS} who proposed an $\mathcal{N} = 8$ supersymmetric field theory based on a Lie 3-algebra. When the generators of the 3-algbera are taken to be finite-dimensional, and the metric positive definite, it was shown in \cite{Gaunt2}, \cite{Papa} that there is essentially one unique example in which the structure constants take the form $f^{abcd} \propto \varepsilon^{abcd}$. In \cite{LandT} and \cite{Mukhi} it was shown that this theory describes two M$2$-branes in an ${\mathbb{R}}^8 / {\mathbb{Z}}_2$ orbifold background. The worldvolume and spacetime superalgebra of this theory has been studied in \cite{Pass}-\cite{IJeon2} (For a discussion of M2-M5 configurations and BPS states in BLG theory and ABJM theory see for example \cite{Park1}-\cite{Thompson}). However it is also possible to define an infinite-dimensional Lie 3-algebra based on the Nambu-Poisson bracket. Remarkably in \cite{M5M2}-\cite{3form} it was shown that the Bagger-Lambert-Gustavsson theory with Nambu-Poisson bracket can be thought of as describing an M5-brane in a three-form background. This can be seen by re-writing the BLG field theory on the 3-dimensional membrane worldvolume $\mathcal{M}$  as a field theory on a 6-dimensional manifold $\mathcal{M} \times \mathcal{N}$ where $\mathcal{N}$ is an internal 3-manifold with Nambu-Poisson structure. The bosonic content of this model was shown to include  a self-dual gauge field on $\mathcal{M} \times \mathcal{N}$, as well as 5 scalar fields parameterising directions transverse to the 6-dimensional worldvolume. In \cite{3form} it was shown that the gauge transformation defined by the Lie 3-algebra associated with the infinite dimensional Nambu-Poisson bracket can be identified as the volume preserving diffeomorphism of the manifold $\mathcal{N}$ on which the 3-algebra acts. Furthermore it was shown  that by performing a double dimensional reduction, the BLG theory with Nambu-Poisson bracket reduces to the Poisson bracket terms of a non-commutative $U(1)$ gauge theory that can be interpreted as the effective worldvolume theory of a D4-brane in a background with a strong NS-NS 2-form gauge field. 

The aim of this paper is to investigate the superalgebra of this theory. Given the interpretation of the Nambu-Poisson model as a description of an M5-brane in 3-form background, one might expect to find central charges corresponding to M5-M2 and M5-M5 intersections in the presence of background flux. Indeed we will see that the central charges appearing in the superalgebra may be interpreted as energy bounds of M5-brane worldvolume solitons in the presence of a background three-form gauge field. In the next section we will briefly review the model proposed in \cite{3form} for the Bagger-Lambert theory with Nambu-Poisson 3-algebra. In section 3 we will re-write the BLG M2-brane charges in terms of Nambu-Poisson variables and derive the central charges of the M5-brane theory. In Section 4 we will look at the double dimensional reduction of this superalgebra and derive the worldvolume superalgebra corresponding to a D4-brane in B-field background. In Section 5 we will discuss our results. An appendix summarises the conventions used throughout the paper.

\section{BLG Theory with Nambu-Poisson Bracket}
The Nambu-Poisson bracket may be regarded as the definition of a Lie 3-algebra in the infinite dimensional space $C (\mathcal{N})$. The basis of $C(\mathcal{N})$ is defined by $\chi^a (y)$ with $a = 1, \ldots \infty$. The Lie 3-algebra structure constant is defined by the Nambu-Poisson bracket as
\begin{equation}
\{\chi^a, \chi^b, \chi^c \} = \sum_{\mud \nud \lambdad} \e_{\mud \nud \lambdad} \partial_{\mud} \chi^a \partial_{\nud} \chi^b \partial_{\lambdad} \chi^c = \sum_{d} f^{abc}_{\ \ \ d} \chi^d (y). \label{a}
\end{equation}
The inner product may be defined as integration over the manifold $\mathcal{N}$ 
\begin{equation}
\langle f, g \rangle = \frac{1}{g^2} \int_{\mathcal{N}} d^3 y f(y) g(y).
\end{equation}
The inner product between basis elements defines the metric
\begin{equation}
h^{ab} = \langle \chi^a , \chi^b \rangle.
\end{equation}
The basic idea of \cite{M5M2} was to expand the 3 dimensional fields appearing in the BLG theory in terms of the basis generators $\{ \chi^a (y) \}$. This results in 6-dimensional fields defined on $\mathcal{M} \times \mathcal{N}$. Of particular importance is the BLG gauge field which may be expressed as a bi-local field defined by
\begin{equation}
A_{\mu} (x, y ,y') = A_\mu^{ab} (x) \chi^a (y) \chi^b (y'). \label{b}
\end{equation}
Taylor expanding the left hand side of this expression around $\Delta y^{\mud} = y^{' \mud} - y^{\mud}$ results in
\begin{equation}
A_{\mu} (x, y, y') = a_\mu (x,y) + b_{\lambda \mud } (x,y) \Delta y^{\mud} + \frac{1}{2} c_{\lambda \mud \nud} (x,y) \Delta y^{\mud} \Delta y^{\nud} + \ldots
\end{equation}
where 
\begin{equation}
b_{\lambda \mud} (x, y) = \frac{\partial}{\partial y^{'\mud}} A_{\lambda} (x, y ,y') |_{y'=y} \label{c}
\end{equation}
It turns out that this is the only component in the Taylor expansion that ever appears in the Bagger-Lambert action. This follows from the fact that $A_{\lambda ab}$ always appears in the action in the form $f^{bcd}_{\ \ \ a}A_{\lambda bc}$ and using \eqref{a}, \eqref{b} and \eqref{c}
\begin{align}
f^{bcd}_{\ \ \ a} A_{\lambda bc} &= \e^{\mud \nud \rhod} \langle \partial_{\mud} b_{\lambda \nud} \partial_{\rhod} \chi^d , \chi_a \rangle.
\end{align}
So for example the covariant derivative can be written as
\begin{align}
\D_\lambda X^I (x,y) &\equiv [\partial_\lambda X^{Ia} (x) - g f^{bcd}_{\ \ \ a} A_{\lambda bc} X^I_d (x)] \chi^a (y) \nn \\
&= \partial_\lambda X^I (x,y) - g \e^{\mud \nud \rhod} \partial_{\mud} b_{\lambda \nud} (x,y)\partial_{\rhod} X^I (x,y) \nn \\
&= \partial_\lambda X^I - g \{ b_{\lambda \nud} , y^{\nud} , X^I \}
\end{align}
We see that the only term from the Taylor expansion that appears in the covariant derivative is the two-form gauge field $b_{\lambda \nud}$. In \cite{3form} this two-form gauge field is interpreted as the gauge potential associated with volume preserving diffeomorphisms of $\mathcal{N}$. It is possible to define another type of covariant derivative $\D_{\mud}$ by observing that, for a scalar field $\Phi$, the object $\{X^{\nud}, X^{\rhod}, \Phi \}$ transforms covariantly under volume preserving diffeomorphisms. Thus we can define $\D_{\mud}$ as
\begin{align}
\D_{\mud} \Phi &= \frac{g^2}{2} \e_{\mud \nud \rhod} \{ X^{\nud} , X^{\rhod}, \Phi \} \nn \\
&= \partial_{\mud} \Phi + g (\partial_{\lambdad} b^{\lambdad} \partial_{\mud} \Phi- \partial_{\mud} b^{\lambdad} \partial_{\lambdad} \Phi) + \frac{g^2}{2} \e_{\mud \nud \rhod} \{ b^{\nud} , b^{\rhod}, \Phi \}
\end{align}
where the field $b^{\mud}$ is defined through the relation
\begin{equation}
X^{\mud} = \frac{1}{g}y^{\mud} + b^{\mud}, \quad \quad b_{\mud \nud} = \e_{\mud \nud \rhod} b^{\rhod}.
\end{equation}
This may be interpreted as a weakened form of the static gauge condition relating to the fact that $y^{\mud}$ space does not possess full diffeomorphism invariance but only volume preserving diffeomorphism invariance. This implies that one cannot completely fix the $X^{\mud}$ fields, resulting in residual degrees of freedom parameterised by $b^{\mud}$. The quantity $X^{\mud}$ defined above can be shown to transform as a scalar under volume preserving diffeomorphisms. This fact allows one to construct gauge field strengths which transform as scalars under gauge variations. We now have two types of covariant derivative, $\D_{\mu}$ and $\D_{\mud}$ which together constitute a set of derivatives acting on the 6-dimensional space $\mathcal{N} \times \mathcal{M}$. As a special case of these covariant derivatives one may define the following field strengths
\begin{align}
\mH_{\lambda \mud \nud} &= \e_{\mud \nud \lambdad} \D_{\lambda} X^{\lambdad} \\ \nn 
&= H_{\lambda \mud \nud} - g \e^{\rhod \sigmad \lambdad} \partial_{\rhod} b_{\lambda \sigmad} \partial_{\rhod} b_{\mud \nud} \nn \\
\mH_{\dot{1} \dot{2} \dot{3}} &= g^2 \{ X^{\dot{1}}, X^{\dot{2}}, X^{\dot{3}} \} - \frac{1}{g} = \frac{1}{g} (V-1) \nn \\
&= H_{\dot{1} \dot{2} \dot{3}} + \frac{g}{2} (\partial_{\mud} b^{\mud} \partial_{\nud} b^{\nud} - \partial_{\mud} b^{\nud} \partial_{\nud} b^{\mud} ) + g^{2} \{b^{\dot{1}} , b^{\dot{2}} , b^{\dot{3}} \}
\end{align}
where V is the induced volume defined by
\begin{equation}
V = g^3 \{X^{\dot{1}}, X^{\dot{2}}, X^{\dot{3}} \}.
\end{equation}
In \cite{3form}, these were the only components of the field strength ${\mathcal{H}}_{\bar{\mu} \bar{\nu} \bar{\rho}}$ ($\bar{\mu} = 0, \ldots 5$) appearing in the action derived from the BLG theory. In the absence of scalar and fermion matter fields, the non-linear chiral field action of \cite{3form} took the form
\begin{equation}
S = - \int d^3x d^3y \left( \frac{1}{4} \mH_{\mu \nud \rhod} \mH^{\mu \nud \rhod} + \frac{1}{12} \mH_{\mud \nud \rhod} \mH^{\mud \nud \rhod} + \frac{1}{2} \e^{\mu \nu \rho} B_{\mu}^{\ \mud} \partial_{\nu} b_{\rho \mud} + g \textrm{ det} B_{\mu}^{\ \mud}\right)
\end{equation} 
where $B_{\mu}^{\ \mud} = \e^{\nud \rhod \mud} \partial_{\nud} b_{\mu \rhod}$. This action is invariant under volume preserving diffeomorphisms but does not possess gauge covariance due to the last two terms (which originate from the BLG Chern-Simons term). In \cite{3form} it was assumed that the field strength components ${\mathcal{H}}_{\mu \nu \rho}$ and ${\mathcal{H}}_{\mu \nu \rhod}$ are dual to $\mH_{\mud \nud \rhod}$ and $\mH_{\mu \nud \rhod}$ respectively. In other words
\begin{equation}
\mH_{\mu \nud \rhod} = \frac{1}{2} \e_{\mu \nu \rho} \e_{\nud \rhod \mud} \mH^{\nu \rho \mud}, \quad \mH_{\mud \nud \rhod} = -\frac{1}{6} \e_{\mud \nud \rhod} \e^{\mu \nu \rho} \mH_{\mu \nu \rho}. \label{birt}
\end{equation}
This was confirmed recently in \cite{Pasti} where it was shown that solving the field equations associated with $b_{\mu \nud}$ and $b_{\mud \nud}$ is tantamount to imposing the Hodge self-duality condition on the non-linear field strength ${\mathcal{H}}_{\bar{\mu} \bar{\nu} \bar{\rho}}$. This allowed the authors to re-write the gauge-field Lagrangian in a gauge-covariant form as
\begin{equation}
S = - \int d^3x d^3 y \left(  \frac{1}{8} \mH_{\mu \nud \rhod} \mH^{\mu \nud \rhod} + \frac{1}{12} \mH_{\mud \nud \rhod} \mH^{\mud \nud \rhod} - \frac{1}{144} \e^{\mu \nu \rho} \e^{\mud \nud \rhod} \mH_{\mu \nu \rho} \mH_{\mud \nud \rhod} - \frac{1}{12g} \e^{\mu \nu \rho} \mH_{\mu \nu \rho} \right).
\end{equation}
The last term in this expression can be interpreted as a coupling of the M5-brane to the constant background $C_3$ field which has non-zero components $C_{\mud \nud \rhod} = \frac{1}{g} \e_{\mud \nud \rhod}$. Following \cite{Pasti}, it is possible to re-write this as
\begin{equation}
\int d^3 x d^3 y \frac{1}{12g} \e^{\mu \nu \rho} \mH_{\mu \nu \rho} = \frac{1}{2} \int \mH_3 \wedge C_3. \label{couple}
\end{equation}
This action possesses full volume preserving diffeomorphism invariance. However the Lorentz symmetry is broken by the presence of the three-form field. This concludes the brief overview of the Nambu-Poisson M5-brane model.

\section{Worldvolume Superalgebra of BLG Theory with Nambu-Poisson Bracket}
In this section we would like to compute the worldvolume superalgebra associated with the BLG theory with Nambu-Poisson bracket. The usual method for calculating the superalgebra of a supersymmetric field theory goes as follows: Firstly one derives the conserved supercurrent, which is the Noether current associated with supersymmetry transformations. The supercharge is then defined by the spatial integral of the zeroth component of the supercurrent. Using the fact that the supercharge is the generator of supersymmetry transformations and that the infinitesimal variation of an anticommuting field is given by $\delta \Phi \propto \{ Q, \Phi \} $, it follows that $\int  \delta J^0 \propto  \{Q , Q \}$. We could in practice follow this procedure for the BLG theory with Nambu-Poisson bracket. However a simpler method involves making use of the superalgbera of multiple M2-branes. Our plan is to take the central charges of the Bagger-Lambert M2-brane superalgebra first  calculated in \cite{Pass} (See also \cite{Kazu}, \cite{IJeon2}), and re-express the 3-dimensional fields in terms of 6-dimensional fields using the conventions of the previous section. We will see that in doing so, the M2-brane central charges will recombine to form M5-brane central charges corresponding to solitons of the worldvolume theory. We will also look at the Bogomoly'ni completion of the Hamiltonian and derive the BPS equations corresponding to the self-dual string soliton and the 3-brane vortex. To begin with we will calculate the Nambu-Poisson BLG supercharge. This will highlight the methodology and introduce useful notation and conventions.

\subsection{Nambu-Poisson BLG Supercharge}
We begin by deriving an expression for the supercharge of the Nambu-Poisson BLG theory. The basic idea is to take the original BLG M2-brane supercharge and expand the fields in terms of the basis $\{\chi^a \}$. As we noted above, the spatial integral of the zeroth component of the supercurrent represents the supercharge, and the worldvolume supercurrent takes the simple form \footnote{For details of the supersymmetry transformations of the BLG theory see \cite{BL2}.}
\begin{equation}
-\be J^{\mu} = \bp^a \g^\mu \delta \psi_a \label{alpha}
\end{equation}
where $\delta \psi_a$ refers to the supersymmetric variation of the fermion field. This is related to the fact that the R-symmetry current $\bp \g^\mu \psi$ and supercurrent $J^\mu$ live in the same supermultiplet, and therefore should be related through a supersymmetry transformation. Evaluating \eqref{alpha} using the original BLG supersymmetry transformation one finds
\begin{equation}
\be J^\mu = - \be \Tr (D_\nu X^I, \g^\nu \g^I \g^\mu \psi )  -  \frac{1}{6} \be  \g^{IJK} \g^{\mu} \Tr ( \{ X^I, X^J, X^K \}, \psi ).
\end{equation}
Here the trace defines an inner product. We wish to re-express this in terms of the 6-dimensional covariant derivatives and field strengths introduced in the previous section. This can be achieved by splitting the scalar field index $I \rightarrow (\mud , i)$ and making the replacements
$[ * , * , * ] \rightarrow g^2 \{ * , * , * \} $ and $\textrm{Tr} \rightarrow \la \ra $. Using this prescription along with the definitions presented in section 2 we can write the supercharge as
\begin{align}
\be Q  = &- \be\g^\nu \g^i \g^0 \la \D_\nu X^i , \psi \ra - \be\g^{\nud} \g^i \g^0 \g_{\dot{1} \dot{2} \dot{3}}\la \D_{\nud} X^i , \psi \ra \nn \\
&- \frac{1}{2} \be \g^\nu \g^{\nud \lambdad} \g^0 \g_{\dot{1} \dot{2} \dot{3}}\la \mH_{\nu \nud \lambdad} , \psi \ra - \be \g_{\dot{1} \dot{2} \dot{3}} \g^0 \la  (\mH_{\dot{1} \dot{2} \dot{3}} + \frac{1}{g}) , \psi \ra \nn \\
&- \frac{g^2}{6} \be \g^{ijk} \g^{0} \la \{ X^i, X^j , X^k \}, \psi \ra - \frac{g^2}{2}\be \g^{\mud} \g^{ij} \g^{0} \la \{X^{\mud}, X^i, X^j \}, \psi \ra \label{ace}
\end{align}
where we have suppressed the integral over the worldvolume coordinates of the M2-brane worldvolume. In deriving this expression we made use of the fact that
\begin{align}
\g_{\mud \nud \rhod} = \g_{\dot{1} \dot{2} \dot{3}} \e_{\mud \nud \rhod}  \quad  \textrm{and} \quad (\g_{\dot{1} \dot{2} \dot{3}})^2 = -1
\end{align}
from which it follows that
\begin{equation}
\e_{\mud \nud \rhod} \g^{\mud \nud} = 2 \g_{\rhod} \g_{\dot{1} \dot{2} \dot{3}}.
\end{equation}
The presence of $\g_{\dot{1} \dot{2} \dot{3}}$ in the second and third terms of \eqref{ace} means that only the SO(1,2)$\times$SO(3) subgroup of the full 6 dimensional Lorentz symmetry is manifest. A similar difficulty was encountered for the fermion kinetic terms in \cite{3form}. There it was shown that it is possible to perform a unitary transformation of the spinor variables 
\begin{equation}
\be = \be' U, \quad \psi = U \psi' \label{qw}
\end{equation}
where U is the matrix
\begin{equation}
U = \frac{1}{\sqrt{2}} (1 - \g_{\dot{1} \dot{2} \dot{3}}).
\end{equation}
Performing this transformation on the supercharge, and using the fact that $[\g^{\mud} , \g_{\dot{1} \dot{2} \dot{3}}] = \{ \g^i , \g_{\dot{1} \dot{2} \dot{3}} \} = \{ \g^\mu , \g_{\dot{1} \dot{2} \dot{3}}\} = 0$ results in
\begin{align}
\be' Q' = &- \be'\g^\nu \g^i \g^0 \la \D_\nu X^i , \psi' \ra - \be' \g^{\nud} \g^i \g^0 \la \D_{\nud} X^i , \psi' \ra \nn \\
&- \frac{1}{2} \be' \g^\nu \g^{\nud \lambdad} \g^0 \la \mH_{\nu \nud \lambdad} , \psi' \ra - \be' \g_{\dot{1} \dot{2} \dot{3}} \g^0 \la  (\mH_{\dot{1} \dot{2} \dot{3}} + \frac{1}{g}) , \psi' \ra \nn \\
&+ \frac{g^2}{6} \be' \g^{ijk} \g^{0} \g_{\dot{1} \dot{2} \dot{3}}\la \{ X^i, X^j , X^k \}, \psi' \ra - \frac{g^2}{2}\be' \g^{\mud} \g^{ij} \g^{0} \la \{X^{\mud}, X^i, X^j \}, \psi' \ra.  \label{ace2}
\end{align}
It is possible to look at the weak coupling limit in which $g \rightarrow 0$. In this limit $\mH \rightarrow H$ and $\D \rightarrow \partial$ and the supercharge becomes\footnote{Note that in deriving the weak coupling limit we have used the additional fermionic shift symmetry introduced in \cite{3form} to eliminate the $1/g$ term from the fermion supersymmetry transformation.}
\begin{equation}
\be'Q' = -\be'  \g^{\bar{\nu}} \g^i \g^0 \la \partial_{\bar{\nu}} X^i , \psi' \ra - \frac{1}{12} \be' \g^{\bar{\mu} \bar{\nu} \bar{\lambda}} \g^0 \la H_{\bar{\mu} \bar{\nu} \bar{\lambda}} , \psi'\ra \label{lol}
\end{equation}
with $\bar{\mu} = (\mu , \mud)$. This expression agrees with the supercharge associated with an abelian $\mathcal{N} = (2,0)$ tensor multiplet.

\subsection{Multiple M2-brane superalgebra}
We would now like to use the method outlined in the previous section to re-express the M2-brane central charges of the original BLG theory. Again this involves expanding the fields in terms of the infinite-dimensional generators $\{ \chi^a \} $ associated with the Nambu-Poisson bracket. Our starting point is the Multiple M2-brane superalgebra as presented in \cite{IJeon2}, which can be written as
\begin{align}
\{ Q, Q \} = &-2 P_\mu \g^\mu \g^0  +  Z_{IJ} \g^{IJ} \g^0  +  Z_{\alpha IJKL} \g^{IJKL} \g^\alpha \g^0  \nn \\
&+  Z_{IJKL} \g^{IJKL}  \label{algebra}
\end{align} 
with central charges
\begin{align}
Z_{IJ} &= - \int d^2 x \Tr (D_\alpha X^I D_\beta X^J \e^{\alpha \beta} - D_0 X^K [X^I, X^J, X^K]) \nn \\
Z_{\alpha IJKL} &= \frac{1}{3} \int d^2 x \Tr (D_\beta X^{I} [X^J, X^K, X^{L}] \e^{\alpha \beta}) \nn \\
Z_{IJKL} &= \frac{1}{4} \int d^2 x \Tr ([X^M, X^I, X^J][X^M, X^K, X^L]). \label{charges}
\end{align}
The integrals are over the spatial coordinates of the M2-brane worldvolume and the trace defines the inner product in terms of the 3-algebra generators. It is possible to use the Bagger-Lambert equations of motion to re-write $Z_{IJ}$ and $Z_{\alpha IJKL}$ as two surface integrals. In \cite{Pass} it was shown that these topological terms correspond to the charges of M2-M2 and multiple M2-M5 intersections (Basu-Harvey solution). The $Z_{IJKL}$ central charge terms were first considered in \cite{Kazu} where the distinction was made between so-called trace elements and non-trace elements\footnote{In \cite{Kazu} trace elements are defined as elements of a linear vector space with an inner product whereas non-trace elements are without a notion of inner-product.}. If one only considers trace elements then $Z_{IJKL}$ vanishes  as a result of the total antisymmetry in $I, J, K, L$ indices and the fundamental identity of the BLG theory.  In \cite{Kazu} it was pointed out that constant background configurations of $X^I$'s which take values in non-trace elements should give rise to BPS brane charges. We will see that the contribution from $Z_{IJKL}$ is crucial in obtaining the M5-brane superalgebra as well as the D4-brane superalgbera.

\subsection{Re-writing of M2-brane central charges}
We begin by by splitting the scalar field index $I \rightarrow (\mud , i)$ and making the replacements
$[ * , * , * ] \rightarrow g^2 \{ * , * , * \} $ and $\textrm{Tr} \rightarrow \la \ra $. In what follows, for the sake of clarity, we will suppress the angle brackets $\langle \rangle$ which denote the inner product. Making repeated use of the conventions outlined in Section 2 and the Appendix results in
\begin{align}
-D_\alpha X^I D_\beta X^J \e^{\alpha \beta} \g^{IJ} \g^0 = &- \D_\alpha X^i \D_\beta X^j \g^{\alpha \beta} \g^{ij}  + \frac{1}{2} \mH_{\alpha \rhod \lambdad} \mH_{\beta \sigmad \omegad} \g^{\rhod} \g^{\lambdad \sigmad \omegad} \g^{\alpha \beta} \nn \\
 &+ \D_\beta X^i \mH_{\alpha \nud \lambdad} \g^{\beta \alpha \nud \lambdad} \g^i \g_{\dot{1} \dot{2} \dot{3}}. \label{beast1} \\
 \nn \\
D_0 X^I \{ X^I, X^J , X^K \} \g^{JK} \g^0 = &- 2 \D_0 X^i \D_{\lambdad} X^i \g^{\lambdad}\g^0 \g_{\dot{1} \dot{2} \dot{3}} + 2 \D_{\omegad} X^i \mH_{0 \nud \omegad}  \g^{\nud} \g^i \g^0 \nn \\
&+ \frac{g^2}{2} \mH_{0 \sigmad \lambdad} \{ X^{\mud} , X^{\nud} , X^{\rhod} \} \e^{\mud \sigmad \lambdad} \g^{\nud \rhod} \g^0 + \frac{g^2}{2} \mH_{0 \rhod \lambdad} \{ X^{\mud}, X^j , X^k \} \g^{\mud \rhod \lambdad} \g^{jk} \g^0 \g_{\dot{1} \dot{2} \dot{3}} \nn \\
&+ g^2D_0 X^i \{ X^i,  X^{\nud} , X^j \} \g^{\nud} \g^j \g^0 + g^2 \D_0 X^i \{ X^i , X^j , X^k \} \g^{jk} \g^0. \label{beast2} \\
\nn \\
Z_{\alpha IJKL} \g^{IJKL} \g^{\alpha 0} = &- \D_{\sigmad} X^i \mH_{\beta \lambdad \omegad} \g^{\sigmad \beta \lambdad \omegad} \g^i + 2 \D_\beta X^i (\mH_{\dot{1} \dot{2} \dot{3}} + \frac{1}{g}) \g^\beta \g^i \g_{\dot{1} \dot{2} \dot{3}}  \nn \\
&+ 2 \D_\beta X^i \D_{\lambdad} X^j \g^{\beta \lambdad} \g^{ij} \g_{\dot{1} \dot{2} \dot{3}} - g^2 \mH_{\beta \rhod \nud} \{ X^{\nud} , X^i , X^j \} \g^{\beta \rhod} \g^{ij} \g_{\dot{1} \dot{2} \dot{3}} \nn \\
&- g^2 \D_\beta X^i \{ X^{\mud} , X^j , X^k \} \g^{\beta \mud} \g^{ijk} + \frac{g^2}{3} \D_\beta X^i \{ X^j, X^k ,X^l \} \g^\beta \g^{ijkl} \nn \\
&+ \frac{g^2}{6} \mH_{\beta \rhod \lambdad} \{ X^i , X^j , X^k \} \g^{\beta \rhod \lambdad} \g^{ijk} \g_{\dot{1} \dot{2} \dot{3}}. \label{beast3} \\
\nn \\
Z_{IJKL} \g^{IJKL} = &- \D_{\lambdad} X^i D_{\nud} X^j \g^{\lambdad \nud} \g^{ij} + g^2 \D_{\rhod} X^j \{ X^{\mud} , X^k , X^l\} \g^{\rhod \mud} \g^{jkl} \g_{\dot{1} \dot{2} \dot{3}} \nn \\
&+ g^2 \D_{\lambdad} X^i \{X^i , X^j, X^k  \} \g^{\lambdad} \g^{jk} \g_{\dot{1} \dot{2} \dot{3}} + 2g^2 \D_{\mud} X^i \{ X^{\mud}, X^i , X^j \} \g^j \g_{\dot{1} \dot{2} \dot{3}} \nn \\
&+\frac{g^4}{2} \{X^{\mud} , X^{\nud} , X^{\rhod} \} \{ X^{\mud} , X^i , X^j \} \g^{\nud \rhod} \g^{ij} + \frac{g^4}{4} \{ X^i , X^j , X^k \} \{ X^i , X^l, X^m \} \g^{jklm} \nn \\
& + \frac{g^4}{4}  \{X^{\mud} , X^i , X^j \} \{ X^{\mud} , X^k , X^l \} \g^{ijkl} - g^4 \{ X^{\mud}, X^i , X^j\} \{ X^i , X^k , X^l \} \g^{\mud} \g^{jkl}  \nn \\
&- g^4 \{X^i , X^j , X^{\mud} \} \{ X^i , X^k , X^{\nud} \} \g^{\mud \nud} \g^{jk} \label{beast4}
\end{align}
Looking at these terms we notice that many of them share the same structure despite originating from different M2-brane charges. The hope is that these terms will combine to form central charges of the M5-brane worldvolume theory. However as it stands, many terms sharing the same structure in \eqref{beast1}-\eqref{beast4} have the wrong relative sign so they are unable to combine. It turns out that this problem can be resolved by multiplying the central charges from the left and right by the unitary matrix U. This is reasonable thing to do since, as mentioned earlier, the method for calculating the M2-brane superalgebra involves calculating $\be \delta J^0 = \be \{ Q , Q \} \e$. We have  seen that when calculating spinor quantities in terms of Nambu-Poisson brackets it is necessary to perform unitary transformations on the spinor fields to bring them to the correct form (for example the fermion kinetic terms, fermion supersymmetry transformation and supercharge).  Therefore one might expect that the correct expression for the M5-brane superalgbera would result from evaluating $\be' \{ Q' , Q' \} \e'$. This is equivalent to multiplying the central charges in \eqref{beast1}-\eqref{beast4} from the left and right by the unitary matrix $U$. We will see in the next section that by performing this unitary transformation, all terms of similar structure combine into M5-brane charges.

\subsection{Nambu-Poisson BLG Superalgebra}
It is well known that the M5-brane worldvolume theory contains 1/2 supersymmetric solutions corresponding to worldvolume solitons. A remarkable feature of these soliton solutions is that they suggest their own 11-dimensional spacetime interpretation in terms of 1/4 supersymmetric intersecting brane configurations. It was shown in \cite{Berg} that the spacetime interpretation is already implicit at the level of the worldvolume supersymmetry algebra. There it was shown that the $D=6$  worldvolume supersymmetry algebra of the M5-brane contains two types of central charge
\begin{align}
W &= \int dX \wedge H \quad \quad Z =  \int dX \wedge d Y. \label{Mcharge}
\end{align}
The W-charge is a worldvolume 1-form carried by the self-dual string living on the M5-brane. The $Z$-charge is a worldvolume self-dual 3-form carried by worldvolume 3-branes corresponding to an M5-M5 intersection. $H$ is the worldvolume 3-form field strength and $X$ and $Y$ are two worldvolume scalars describing fluctuations in two directions transverse to the M5-brane worldvolume in spacetime. The integrals are over the subspaces of the 5-dimensional worldspace transverse to the p-brane solitons. $W$ dependends on one scalar field, as required by its identification with the magnitude of a 1-form in the space transverse to the brane. Similarly, the dependence of the 3-brane charge Z on two scalars is required by its interpretation as a transverse 2-form. In this section we will combine the terms appearing in the previous section into M5-brane central charges and write the BLG Nambu-Poisson superalgebra. At lowest order in the coupling $g$ we will see that the superalgebra contains both the $W$ and $Z$ charges expected from the worldvolume superalgebra of the M5-brane. In addition we will find higher order coupling central charges, some of which involve the background 3-form $C$-field. Since the resulting expression is long and complicated we will split the superalgebra into three parts depending on the order of the coupling of the central charge. Having performed the unitary transformation described in the previous section one finds
\begin{equation}
\{ Q, Q \}_{\textrm{central}} \propto  Z (g^0)\g_{\dot{1} \dot{2} \dot{3}} + Z(g^2) \g_{\dot{1} \dot{2} \dot{3}} + Z(g^4) \g_{\dot{1} \dot{2} \dot{3}}
\end{equation}
We will take each term separately and try to provide an interpretation of the central charges. To begin with we look at $Z(g^0)$. Remarkably one finds the compact expression 
\begin{align}
Z(g^0) = &+ \D_a X^i \D_b X^j \g^{ab} \g^{ij} + 2 \D_{0} X^i D_{\lambdad} X^i \g^0 \g^{\lambdad} + \frac{1}{3}\D_a X^i \mH_{bcd} \g^{abcd} \g^i \nn \\
 &-\frac{2}{g} \D_{\beta} X^i \g^{\beta} \g^i \g_{\dot{1} \dot{2} \dot{3}} + \mH_{0 \mud \nud} (\mH^{\mud \nud \rhod} + \frac{1}{g} \e^{\mud \nud \rhod}) \g^0 \g^{\rhod} -\frac{1}{2} \mH_{\alpha \rhod \lambdad} \mH_{\beta \sigmad \omegad} \g^{\rhod} \g^{\lambdad \sigmad \omegad} \g^{\alpha \beta} \nn
\end{align}
with $a = (\alpha , \mud)$ representing the spatial coordinates on the worldvolume of the M5-brane. Note that in obtaining this result we made use of $\mH^{\alpha \beta \lambdad} = - \frac{1}{2} \e^{\alpha \beta} \e^{\nud \rhod \lambdad} \mH_{0 \nud \rhod}$ as well as 
\begin{align}
\frac{1}{3}\D_a X^i \mH_{bcd} \g^{abcd} = &+ \D_\alpha X^i \mH_{\beta \mud \nud} \g^{\alpha \beta \mud \nud}+  \D_{\mud} X^i \mH_{\beta \nud \lambdad} \g^{\mud \beta \nud \lambdad} \nn \\
&+  \D_{\mud} X^i \mH_{\beta \gamma \nud} \g^{\mud \beta \gamma \nud} + \frac{1}{3}\D_\alpha X^i \mH_{\nud \sigmad \lambdad} \g^{\alpha \nud \sigmad \lambdad}.
\end{align} 
We see that there are three types of term in $Z(g^0)$: Charges of the form $\D X \D X$, those of the form $\D X \mH$ and finally those of the form $\mH \mH$. We will only be interested in static configurations so will ignore the term involving $\D_0X$. We see that the first term $\D_a X^i \D_b X^j$ corresponds to the charge of the 3-brane vortex living on the M5-brane worldvolume. When only two scalar fields are active ($X$ and $Y$) we can identify it as the charge $Z$ appearing in \eqref{Mcharge}. The $\D \mH$ term corresponds to the self-dual string charge. If we consider the situation in which only one scalar field is active ($X$) and assume that this scalar is a function of only four of the spatial worldvolume coordinates of the M5 brane, namely $\hat{\mu} = (2, \dot{1}, \dot{2}, \dot{3})$, we see that the $\D \mH$ term becomes $\e^{\hat{\mu} \hat{\nu} \hat{\rho} \hat{\sigma}} \D_{\hat{\mu}} X  \mH_{\hat{\nu} \hat{\rho} \hat{\sigma}}$. This exactly corresponds to the energy bound $W$ appearing in \eqref{Mcharge}. The  $\D \mH$ term with coefficient $2/g$ may be thought of as a contribution from the background 3-form gauge field $C$. Making the identification $C_{\dot{1} \dot{2} \dot{3}} = \frac{1}{6} \e^{\mud \nud \rhod} C_{\mud \nud \rhod} \propto 1/g$ we have  
\begin{equation}
-\frac{2}{g} \D_{\beta} X^i \g^{\beta} \g^i \g_{\dot{1} \dot{2} \dot{3}} = \frac{1}{3} \D_\beta X^i C_{\mud \nud \rhod} \g^{\beta \mud \nud \rhod} \g^i.
\end{equation}
We can think of this term as representing a C-field modification of the self-dual string charge in the directions $\dot{1} \dot{2} \dot{3}$. We can provide an interpretation of the $\mH \mH$ term by thinking about its double dimensional reduction. This will be carried out explicitly in the next section but for now we note that the $\mH_{\alpha \rhod \lambdad} \mH_{\beta \sigmad \omegad}$ charge reduces to a D4-brane instanton charge that looks like $F\tilde{F}$. From the D4-brane perspective this can be thought of as the charge of a D0-brane within the worldvolume of the D4-brane. Thus from the M5-brane perspective this charge can be thought of as an M-wave intersecting an M5-brane. We now turn to the $Z(g^2)$ charges,
\begin{align}
Z(g^2) = &+ g^2 \D_a X^i \{ X^{\mud} , X^j , X^k \} \g^{a \mud} \g^{ijk} - g^2 D_{\bar{a}} X^i \{ X^i , X^j, X^k \} \g^{\bar{a}} \g^{jk} \g_{\dot{1} \dot{2} \dot{3}} \nn \\
&- \frac{g^2}{3} \D_\beta X^i \{ X^j , X^k , X^l \} \g^\beta \g^{ijkl} \g_{\dot{1} \dot{2} \dot{3}} + 2g^2 \D_{\mud} X^i \{ X^{\mud} , X^i , X^j \} \g^j \nn \\
&+ g^2 \D_0 X^i \{ X^{\nud} , X^i , X^j \} \g^0 \g^{\nud} \g^j + \frac{g^2}{2} \mH_{0 \rhod \lambdad} \{X^{\mud} , X^j , X^k \} \g^{\mud \rhod \lambdad} \g^{jk} \g^0 \nn \\
&- g^2 \mH_{\beta \rhod \nud} \{ X^{\nud} , X^i , X^j \} \g^{\beta \rhod} \g^{ij} - \frac{g^2}{6} \mH_{\beta \rhod \lambdad} \{ X^i , X^j , X^k \}\g^{\beta \rhod \lambdad} \g^{ijk} \nn \\
&-2g^2 (\mH_{\dot{1} \dot{2} \dot{3}} + \frac{1}{g}) \{ X^{\mud}, X^i , X^j \} \g^{\mud} \g^{ij}. \label{Basu1}
\end{align}
In the second term the label $\bar{a} = (0 , \mud)$. A few comments are in order. As we will see in the next section, double dimensional reduction of the first term gives rise to a charge $g\tD_a X^i \{ X^j, X^k \}$. 
This term bares a structural similarity to the charge corresponding to a Nahm equation configuration in which multiple D2-branes intersect a D4-brane in which the commutators have been replaced by Poisson-brackets. Thus one might expect to find a Nambu-Poisson bracket version of the Basu-Harvey type energy bound in the C-field modified M5-brane superalgebra. Indeed the charge $\frac{g^2}{3} \D_\beta X^i \{ X^j , X^k , X^l \}$ appearing in \eqref{Basu1} is reminiscent of the Basu-Harvey charge expressed in terms of Nambu-Poisson bracket. 
For a discussion of the geometry of the M5-brane in the presence of a constant C-field as well as the derivation of the C-field modified Basu-Harvey equation as a boundary condition of the multiple M2-brane theory see \cite{Douglas}. 
Finally we have the $Z(g^4)$ charges which take the form
\begin{align}
Z(g^4) = &+ g^4 \{ X^i , X^j , X^{\mud}\} \{ X^i , X^k , X^{\nud} \} \g^{\mud \nud} \g^{jk} + g^4 \{X^{\mud} , X^i , X^j \} \{ X^i , X^k , X^l \} \g^{\mud} \g^{jkl} \g_{\dot{1} \dot{2} \dot{3}}  \nn \\
&- \frac{g^4}{4} \{ X^{\mud} , X^i , X^j \} \{ X^{\mud} , X^k , X^l\} \g^{ijkl} - \frac{g^4}{4} \{X^i , X^j , X^k \} \{ X^i , X^l , X^m \} \g^{jklm}. 
\end{align}
We will see that only the first and third terms survive the dimensional reduction and give rise to a charge analogous to the D4-brane charge found in Matrix theory. The M5-brane analogue of this interpretation requires further investigation. Most importantly in this section we have seen that the BLG superalgebra based on Nambu-Poisson bracket contains the expected central charges corresponding to an M5-brane. Namely the 3-brane charge and self-dual string charge, suitably modified by the presence of the background gauge field. In the next section we will confirm the existence of these charges as energy bounds from the Hamiltonian perspective.

\subsection{Hamiltonian analysis and BPS equations}
We would like to derive the energy bounds corresponding to M5-M2 and M5-M5 intersections by looking at the Bogomoly'ni completion of the Nambu-Poisson BLG Hamiltonian. For the original Bagger-Lambert theory the energy-momentum tensor with fermions set to zero takes the form
\begin{equation}
T_{\mu \nu} = D_\mu X^I D_\nu X^I - \eta_{\mu \nu} (\frac{1}{2} D_\rho X^I D^\rho X^I + V).
\end{equation}
The Chern-Simons term does not contribute due to the fact that it is topological in nature and does not depend on the worldvolume metric. For static configurations the energy density takes the form
\begin{equation}
E = \frac{1}{2} D_\alpha X^I D_\alpha X^I + \frac{1}{12} [X^I , X^J , X^K]^2.
\end{equation}
Re-writing this expression in terms of M5-brane fields one finds
\begin{align}
E = &+\frac{1}{2}(\D_{\alpha} X^i)^2 + \frac{1}{2} (\D_{\mud} X^i)^2 + \frac{1}{4} \mH^2_{\alpha \mud \nud} + \frac{1}{2}(\mH_{\dot{1} \dot{2} \dot{3}} + C_{\dot{1} \dot{2} \dot{3}})^2 \nn \\
&+ \frac{g^4}{4} \{ X^{\mud} , X^i , X^j \}^2 + \frac{g^4}{12} \{ X^i , X^j , X^k \}^2 . \label{energy}
\end{align}
In order to find the energy bound corresponding to the self-dual string soliton we consider the situation in which there is only one active scalar field which we call $X$. Furthermore we assume that this scalar is a function of only four of the spatial worldvolume coordinates of the M5 brane, namely $\hat{\mu} = (2, \dot{1}, \dot{2}, \dot{3})$. In what follows we will assume $C_{\dot{1} \dot{2} \dot{3}} = 0$. In this case the energy density takes the following form
\begin{equation}
E = \frac{1}{2}(\D_{\hat{\mu}} X)^2 + \frac{1}{12} \mH^2_{\hat{\mu} \hat{\nu} \hat{\rho}} 
\end{equation}
We can re-write this as 
\begin{equation}
E = \frac{1}{2}|\D_{\hat{\mu}} X \pm \frac{1}{6}\e_{\hat{\mu}}^{\ \hat{\nu} \hat{\rho} \hat{\sigma}} \mH_{\hat{\nu} \hat{\rho} \hat{\sigma}}|^2 \mp \frac{1}{6}\e^{\hat{\mu} \hat{\nu} \hat{\rho} \hat{\sigma}} \D_{\hat{\mu}} X  \mH_{\hat{\nu} \hat{\rho} \hat{\sigma}} 
\end{equation}
We see that the energy density is minimised when the BPS equation
\begin{equation}
\D_{\hat{\mu}} X \pm \frac{1}{6}\e_{\hat{\mu}}^{\ \hat{\nu} \hat{\rho} \hat{\sigma}} \mH_{\hat{\nu} \hat{\rho} \hat{\sigma}} = 0 \label{cunt}
\end{equation}
is satisfied. In this case the energy is bounded by the central charge
\begin{equation}
Z = \mp \frac{1}{6}\e^{\hat{\mu} \hat{\nu} \hat{\rho} \hat{\sigma}} \D_{\hat{\mu}} X  \mH_{\hat{\nu} \hat{\rho} \hat{\sigma}}. \label{cunt2}
\end{equation}
The BPS equation \eqref{cunt} matches the result found recently in \cite{Furu}, where the author used supersymmetry arguments to derive the BPS equation corresponding to the self dual string soliton. Furthermore the energy bound \eqref{cunt2} matches the central charge found in the superalgebra of the previous section. The effect of turning on the $C_{\dot{1} \dot{2} \dot{3}}$ field is to modify the BPS equation and energy bound  by shifting the $\mH_{\dot{1} \dot{2} \dot{3}}$ component of the field strength.  In \cite{3form} a generalisation of the Seiberg-Witten map was shown to relate the theory of M5-branes based on Nambu-Poisson bracket with the theory of an M5-brane in constant 3-form background. In \cite{Furu},  solutions to the BPS equation \eqref{cunt} were found up to first order in the coupling $g$. The generalised Seiberg-Witten map  was then used to match the solution with the known results derived in \cite{Michi}, \cite{Youm}. We have shown that the energy bound corresponding to this solution can be seen from the Hamiltonian and superalgebra perspective. Next we will consider the 3-brane soliton corresponding to an M5-M5 intersection. We will let the 3-brane lie in the $(0, 1, 2, \dot{1})$ plane with transverse directions in the $\dot{a} = (\dot{2}, \dot{3})$ directions. We allow two scalars to be non-zero and label them $X$ and $Y$.  These scalars are only functions of the transverse directions $(\dot{2}, \dot{3})$. In this case the energy density \eqref{energy} becomes
\begin{align}
E &= \frac{1}{2}(\D_{\dot{a}} X)^2  + \frac{1}{2}(\D_{\dot{a}} Y)^2  \nn \\
&= \frac{1}{2}|\D_{\dot{a}} X \pm \e_{\dot{a} \dot{b}}\D_{\dot{b}} Y |^2 \mp \e_{\dot{a} \dot{b}} \D_{\dot{a}} X \D_{\dot{b}} Y
\end{align}
Thus we see that the energy is bounded by 
\begin{equation}
Z = \e_{\dot{a} \dot{b}} \D_{\dot{a}} X \D_{\dot{b}} Y.
\end{equation}
This matches the central charge of the 3-brane vortex found in the superalgebra of the previous section. This concludes our discussion of the M5-brane central charges. In the next section we will investigate the double dimensional reduction of the M5-brane Nambu-Poisson superalgebra.
\newpage
\section{Double Dimensional Reduction of Superalgebra}
In this section we perform a double-dimensional reduction of the Nambu-Poisson M5-brane superalgebra. We then attempt to provide an interpretation of the corresponding charges in terms of spacetime intersections involving the D4-brane. Importantly we will see that the algebra we derive consists of only the `Poisson-bracket' terms of the full non-commutative D4-brane superalgebra.  
\subsection{Dimensional Reduction Conventions}
In order to perform the double-dimensional reduction we choose the compactification direction to be $X^{\dot{3}}$. Following the conventions of \cite{3form} we define the gauge potential as
\begin{equation}
{\hat{a}}_\mu = b_{\mu \dot{3}}, \quad \quad {\hat{a}}_{\dot{\alpha}} = b_{\dot{\alpha} \dot{3}}. 
\end{equation}
As a result the covariant derivatives become
\begin{equation}
D_\mu X^{\dot{\alpha}} = - \e^{\dot{\alpha} \dot{\beta}} {\hat{F}}_{\mu \dot{\beta}}, \quad D_\mu X^{\dot{3}} = - {\tilde{a}}_\mu, \quad D_\mu X^i = {\hat{D}}_\mu X^i,
\end{equation}
where we define
\begin{align}
{\hat{F}}_{ab} &= \partial_a {\hat{a}}_b - \partial_b {\hat{a}}_a +g \{{\hat{a}}_a , {\hat{a}}_b \}, \nn \\
{\tilde{a}}_\mu &= \e^{\dot{\alpha} \dot{\beta}} \partial_{\dot{\alpha}} b_{\mu \dot{\beta}}, \nn \\
{\hat{D}}_\mu \Phi &= \partial_\mu \Phi + g \{{\hat{a}}_\mu , \Phi \}.
\end{align}
The Poisson bracket $\{ *, * \}$ is defined as the reduction of the Nambu-Poisson bracket
\begin{equation}
\{f,g \} = \{y^{\dot{3}}, f, g \}.
\end{equation}
The relevant Nambu-Poisson Brackets are
\begin{align}
\{X^{\dot{1}}, X^{\dot{2}}, X^{\dot{3}} \} &= \frac{1}{g^2} {\hat{F}}_{\dot{1} \dot{2}} + \frac{1}{g^3} \label{please} \\
\{ X^{\dot{3}}, X^{\dot{\alpha}} , X^i \} &= \frac{1}{g^2} \e^{\dot{\alpha} \dot{\beta}} {\hat{D}}_{\dot{\beta}} X^i  \\
\{X^{\dot{3}}, X^i , X^j \} &= \frac{1}{g} \{ X^i , X^j \}
\end{align}
To perform the dimensional reduction it will prove easiest if we start with the M2-brane central charges and re-write them in terms of the D4-brane variables directly\footnote{This will give the same result as performing the reduction on the M5-brane superalgebra derived in the previous section.}. 

\subsection{D4-brane Superalgebra}
We begin by using the conventions of the previous section to re-write the central charges of the M2-brane theory. One finds
\begin{align}
-D_\alpha X^I D_\beta X^J \e^{\alpha \beta} \g^{IJ} \g^0 \rightarrow &+ \tD_\alpha X^i \tD_\beta X^j \g^{\alpha \beta} \g^{ij} + 2 \tF_{\betad \alpha} \tD_\beta X^i \g^{\betad \alphad \beta} \g^i \g^{\dot{3}} \nn \\
&+ \tF_{\alpha \gammad} \tF_{\beta \deltad} \g^{\alpha \gammad \beta \deltad} + 2 \tF_{0 \beta} \tF_{\beta \gammad} \g^{0 \gammad} \nn \\ 
&+ 2 \tD_\beta X^i \tF_{0 \beta } \g^{\dot{3} i } \g^0. \nn \\
\nn \\
D_0 X^K \{ X^K, X^I, X^J \} \g^{IJ} \g^0 \rightarrow &+ \frac{1}{2} \tF_{\alpha \beta} \tF_{\alphad \betad} \g^{\alpha \beta \alphad \betad} + \frac{1}{g} \tF_{\alpha \beta} \g^{\dot{1} \dot{2}} \g^{\alpha \beta} \nn \\
&+ \tF_{\alpha \beta} \tD_{\gammad} X^i \g^{\alpha \beta \gammad} \g^i \g^{\dot{3}} + \frac{g}{2} \tF^{\alpha \beta} \{X^i, X^j \} \g^{\alpha \beta} \g^{ij} \nn \\
&+ \tF_{0 \betad} \tF_{\alphad \betad} \g^{0 \betad \alphad \gammad} + 2\tF_{0 \betad} \tF_{\betad \gammad} \g^{0 \gammad} \nn \\
&+ \frac{2}{g} \tF_{0 \betad} \g^{0 \betad} \g^{\dot{1} \dot{2}} + 2 \tD_{\betad} X^i \tF_{0 \betad } \g^{\dot{3} i}\g^0  \nn \\ 
&+ 2g \tD_0 X^i \{ X^i , X^j \} \g^0 \g^j \g^{\dot{3}} - 2\tD_0 X^i \tD_{\betad} X^i \g^{0 \betad} . \nn 
\end{align}
\begin{align}
\frac{1}{3} \tD_\beta X^I \{ X^J, X^K, X^L \} \g^{IJKL} \e^{\alpha \beta} \g^{\alpha 0} \rightarrow &+ \tF_{\alphad \betad} \tD_{\gamma} X^i \g^{\alphad \betad \gamma} \g^i \g^{\dot{3}} + \frac{2}{g} \tD_{\beta} X^i \g^i \g^{\dot{1} \dot{2} \dot{3}} \g^\beta \nn \\
&+ g \tF_{\alpha \betad} \{ X^i, X^j \} \g^{\alpha \betad} \g^{ij} + g \tD_\beta X^i \{ X^j , X^k \} \g^{\dot{3}} \g^{\beta} \g^{ijk} \nn \\
&+ 2\tF_{\alpha \betad} \tD_{\gammad} X^i \g^{\alpha \betad \gammad} \g^i \g^{\dot{3}} + 2 \tD_\alpha X^i \tD_{\betad} X^j \g^{\alpha \betad}. \nn \\
\nn \\
\frac{1}{4} \{X^M, X^I, X^J \} \{ X^M, X^K, X^L \} \g^{IJKL} \rightarrow &+ \frac{g}{2}\tF_{\alphad \betad} \{ X^i , X^j \} \g^{\alphad \betad} \g^{ij} +  \{ X^i , X^j \} \g^{\dot{1} \dot{2}} \g^{ij} \nn \\
&+ g \tD_{\alphad} X^i \{ X^j , X^k \} \g^{\dot{3}} \g^{\betad} \g^{ijk} + \tD_{\alphad} X^i \tD_{\betad} X^j \g^{\alphad \betad} \g^{ij} \nn \\
&+ \frac{g^2}{4} \{ X^i , X^j \} \{ X^k , X^l \} \g^{ijkl}.
\end{align}
Looking at these terms we notice that many of them share the same structure despite originating from different M2-brane charges. Combining these terms it is possible to write the D4-brane superalgbera as
\begin{align}
\{Q , Q \}_{\textrm{central}} = &+ \tD_a X^i \tD_b X^j \g^{ab} \g^{ij}  + \tD_a X^i \tF_{bc}  \g^{abc} \g^i \g^{\dot{3}} - 2 \tD_a X^i \tF_{0a} \g^0 \g^i \g^{\dot{3}}\nn \\
&+ \frac{1}{4}\tF_{ab}\tF_{cd} \g^{abcd} + \frac{g}{2} \tF_{ab} \{ X^i, X^j \} \g^{ab} \g^{ij} + g \tD_a X^i \{ X^j, X^k \}  \g^{a} \g^{ijk} \g^{\dot{3}} \nn \\
&+\frac{g^2}{4} \{ X^i , X^j \} \{ X^k , X^l \} \g^{ijkl} + \frac{1}{g} \tF_{\alpha \beta} \g^{\dot{1} \dot{2}} \g^{\alpha \beta} +\frac{2}{g} \tD_{\beta} X^i \g^i \g^{\dot{1} \dot{2} \dot{3}} \g^\beta \nn \\
&+ \{ X^i , X^j \} \g^{\dot{1} \dot{2}} \g^{ij}+ 2g \tD_0 X^i \{ X^i , X^j \} \g^0 \g^j \g^{\dot{3}} - 2\tD_0 X^i \tD_{\betad} X^i \g^{0 \betad} . \label{eight}
\end{align}
\\
\subsection{Interpretation of charges}
In \cite{3form} (see also \cite{PHo}) it was argued that the Nambu-Poisson BLG theory is a theory of an M5-brane in a strong C-field background. This was motivated by the fact that, upon double dimensional reduction, the theory shares a structural similarity to a non-commutative D4-brane theory. 
It is important however to emphasise a few points relating to this interpretation. Firstly, the double dimensional reduction of the Nambu-Poisson M5-brane only captures the Poisson bracket structure of the non-commutative D4-brane action, but misses all the higher order terms in the $\theta$-expansion of the Moyal bracket. Recall that the non-commutative D4-brane theory is described using the Moyal $*$ product (see for example \cite{Seiberg}). It is possible to expand the Moyal bracket in $\theta$. One finds at lowest order in $\theta$ the Poisson bracket structure,
\begin{equation}
[f, g]_{\textrm{Moyal}} = f * g - g * f = \theta^{ij} \partial_i f \partial_j g + \mathcal{O} (\theta^3).
\end{equation}
where
\begin{equation}
f(x) * g(x) = e^{\frac{i}{2} \theta^{ij} \frac{\partial}{\partial \xi^i} \frac{\partial }{\partial \chi^j}} f(x + \xi ) g(x + \chi )|_{\xi = \chi =0}.
\end{equation}
The fact that the double-dimensional reduction of the Nambu-Poisson theory only captures the lowest order term in the Moyal-bracket means that one is not able to identify this theory as the full non-commutative D4-brane theory. Rather it may be viewed as a Poisson-bracket truncation of the full non-commutative field theory.\footnote{In a recent paper \cite{Chen} the question was asked whether it is possible to deform the Nambu-Poisson M5-brane theory such that its double dimensional reduction gives rise to the full non-commutative Yang-Mills theory to all orders in the non-commutativity parameter $\theta$. They found that there is no way to deform the Nambu-Poisson gauge symmetry such that the full non-commutative gauge symmetry was recovered upon double-dimensional reduction.} Given these consideration, plus the fact that higher order terms in $g$ are considered in this paper, it is important to emphasise that the results derived for the central charges in \eqref{eight} are only valid for the Poisson-bracket D4-brane theory. 
\\
\\
In \eqref{eight} the spatial coordinates of the brane are labeled by $a = (\alpha, \alphad)$ . Given the fact that the D4-brane results from the double dimensional reduction of the M5-brane, one would expect to see remnants of the M5-brane solitons in the worldvolume superalgebra of the D4-brane. Furthermore, the M5-brane couplings to the background 3-form should appear as couplings to the NS-NS 2-form B-field from the D4-brane perspective. As outlined in \cite{Howe}, the self dual sring soliton from the M5-brane worldvolume theory can be reduced to a 0-brane or 1-brane solution on the D4-brane worldvolume depending on whether the string is wrapped or un-wrapped around the compact dimension. In the abelian case the BPS equation corresponding to the 0-brane BPS soliton on the D4-brane takes the form $F_{0a} \propto \partial_a X$ where $X$ is the only active scalar field parameterising a single direction transverse to the brane. From this one can read off the energy bound when the BPS equation is satisfied which is proportional to $ \partial_a X F_{0a}$. We see the covariant generalisation of this term appearing in the superlagebra derived above in the form $D_a X^i F_{0a}$. This bound can naturally be interpreted as the endpoint of a fundamental string on the D4-brane. Alternatively, the 1-brane soliton resulting from the un-wrapped self-dual string has BPS equation $F_{bc} \propto \e_{abc} \partial^a X$ with corresponding energy bound proportional to $ \e_{abc} \partial_a  X F_{bc}$. This term can naturally be compared with $D_a X^i F_{bc}  \g^{abc}$ appearing in \eqref{eight}. One can think of this as the bound associated with a D2-brane intersecting a D4-brane which appears as a 1-brane soliton from the D4-brane perspective. It is worth noting that the D4-brane worldvolume superalgebra allows scalar central charges in the $\bf{1} + \bf{5}$ representation of the $\textrm{Spin}(5)$ R-symmetry group which can be interpreted as the transverse rotation group. As noted above, the scalars in the $\bf{5}$ representation can be interpreted as the endpoints of fundamental strings on the D4-brane. The $\textrm{Spin} (5)$ singlet scalar is a magnetic charge which from the spacetime perspective corresponds to a D0-brane intersecting a D4-brane. Because it is a singlet its charge cannot depend on any transverse scalars. Following arguments similar to those presented in \cite{Gaunt}, one can show that the total energy $E$ of this configuration, relative to the worldvolume vacuum, is subject to the bound $E \ge |Z|$ where Z is the topological charge
\begin{equation}
Z = \frac{1}{4} \int_{\textrm{D4}} \textrm{tr} F \tilde{F}.
\end{equation}
This corresponds to the central charge appearing in the superalgebra
\begin{align}
\frac{1}{4}\tF_{ab}\tF_{cd} \g^{abcd} &\propto \frac{1}{4}\tF_{ab}\tF_{cd} \e^{abcd} \nn \\
&= \frac{1}{4} \tF {\tilde{\tF}}.
\end{align}
The central charge in \eqref{eight} involving two covariant derivatives, $\tD_a X^i \tD_b X^j$, derives from the 3-brane vortex on the M5-brane worldvolume  corresponding to an M5-M5 intersection. The double dimensional reduction of the 3-brane gives rise to a 2-brane on the D4-brane worldvolume. This may be interpreted as the intersection of two D4-branes over a 2-brane. 
It is interesting to notice that the superalgebra contains a central charge term proportional to $\tD_a X^i \{ X^j, X^k \}$. This charge would appear to represent Nambu-Poisson analogue of the energy bound associated with the Nahm equation which describes the intersection of multiple D2-branes with a D4-brane. However it is well known that the Nahm equation is the BPS equation associated with the non-abelian worldvolume theory of D-branes. It therefore seems remarkable that information regarding the energy bound of this configuration should appear in the worldvolume theory of a single D4-brane. Furthermore, the central charge proportional to $\{ X^i, X^j \} \{ X^k, X^l\}$ seems analogous to the D4-brane charge 
appearing in the matrix model for M-theory found in \cite{Banks}. It seems as if the worldvolume superalgebra derived above contains information about intersections involving multiple D0-branes and multiple D2-branes  from the perspective of the D4-brane as well as from the perspective of the D0-brane and D2-brane worldvolume theories.

\subsection{Comments on $1/g$ terms in D4-brane theory}
It is worth making a few comments regarding the interpretation of the $1/g$ terms appearing in the superalgebra \eqref{eight}. In \cite{3form}, the coupling of the dimensionally reduced theory was identified with the non-commutativity parameter of the D4-brane theory, $g = \theta$. Based on this identification one can think of the $1/g$ terms as being $1/ \theta$ modifications of certain central charge terms. More precisely, any term in the algebra \eqref{eight} deriving from the dimensional reduction rule \eqref{please} will contain a term of the form $F_{\dot{1} \dot{2}} + 1/ \theta$ (once we have identified $g = \theta$). 
Ultimately we can think of this shift in the $(\dot{1} \dot{2})$ components of the field strength as being related to the presence of a background B-field. After all, the non-commutative theory is meant to be an effective description of the D4-brane in B-field background. In order to move from the non-commutative description (without background flux) to the `commutative' description involving the background B-field it is necessary to use the Seiberg-Witten map \cite{Seiberg}. 
In principle this can be done for the central charge terms involving $1/\theta$ appearing in \eqref{eight}. It is known for example that from the perspective of the 1-brane soliton on the D4-brane worldvolume, the presence of a background $B$-field  causes a tilting of the 1-brane as it extends from the D4-brane. This can be shown to result from a shift in the field strength by the B-field. This shift leads to a change in the form of the BPS equation and the corresponding BPS energy bound (central charge).\footnote{ There is a slightly puzzling feature of the D4-brane algebra appearing in \eqref{eight}. This is related to the fact that, for the case studied in this paper, it would seem the correct non-commutative description of the D4-brane theory is the matrix model description. This might be inferred from the fact that the Nambu-Poisson M5-brane theory is constructed from the M2-brane action (The analogous situation in type IIA string theory would be constructing the D4-brane from the D2-brane action and we know the matrix theory description naturally arises in this setting). In this case, following \cite{Seiberg2}, one expects to find the combination $F - 1/ \theta$ appearing in the action, not $F + 1/ \theta$. This discrepancy could be related to the fact that the reduction of the Nambu-Poisson algebra only gives the Poisson terms of the non-commutative D4-brane theory, whereas the combination $F- 1 / \theta$ appears in the full non-commutative theory. We hope to return to this issue in a future publication. We would like to thank the referee for bringing this point to our attention.} 
\section{Discussion and Conclusions}
In this paper we investigated the worldvolume superalgebra of the BLG theory based on a Nambu-Poisson bracket. This was achieved by re-expressing the BLG M2-brane worldvolume central charges in terms of infinite dimensional generators defined by the Nambu-Poisson bracket. This resulted in a new set of central charges expressed in terms of 6-dimensional fields. These we interpreted as central charges corresponding to worldvolume solitons of the M5-brane worldvolume theory in the presence of a background 3-form gauge field. In particular we found that the central charges of the M5-brane theory could be grouped according to the power of the coupling constant $g$. At $\mathcal{O} (g^0)$ we found central charges corresponding to the self-dual string soliton which describes an M2-M5 intersection as well as a 3-brane soliton corresponding to an M5-M5 intersection. We also found that background C-field modifies the energy bound of the self-dual string charge by shifting the $\mH_{\dot{1} \dot{2} \dot{3}}$ component of the field strength. Our paper therefore reinforces the results presented in \cite{M5M2} and \cite{3form} in which the BLG theory based on Nambu-Poisson bracket was interpreted as a theory describing an M5-brane in 3-form flux. Furthermore we investigated the double dimensional reduction of the M5-brane Nambu-Poisson superalgebra. We found central charges corresponding to spacetime intersections of the D4-brane with D0-branes, D2-branes and D4-branes, as well as the endpoint of the fundamental string. We also found charges corresponding to Nahm type configurations which we interpreted as multiple D2-branes ending on the D4-brane. Finally we found a charge reminiscent of the D4-brane charge found in Matrix theory. It would appear that the superalgebra of the BLG theory based on Nambu-Poisson structure contains a huge amount of information about spacetime configurations including non-abelian phenomena. In this paper we were only able to provide spacetime interpretations for some of the M5-brane charges. However, knowledge of the D4-brane spacetime configurations may provide a hint. We hope to return to the interpretation of the remaining charges in the near future.
\section*{Acknowledgements\markboth{Acknowledgements}{Acknowledgements}} 
We would like to thank Pei-Ming Ho and Yosuke Imamura for patient explanations of their work. We are also grateful to D. Thompson, D. Berman and B. Spence for useful discussions regarding a draft version of this paper. Finally I would like to thank Nomad Bookshop for their hospitality whilst this work was completed. AL is supported by an STFC grant.

\newpage
\begin{appendix}
\appendix

\section{Conventions}
In the BLG Nambu-Poisson model of the M5-brane, the worldvolume is taken to be the product manifold $\mathcal{M} \times \mathcal{N}$. We assume a Minkowski metric $\eta_{\mu \nu} = \textrm{diag} (-++)$ on $\mathcal{M}$ and a Euclidean metric $\delta_{\mud \nud} = \textrm{diag} (+++)$ on the internal space $\mathcal{N}$. The supersymmetry transformation parameter $\e$ and the fermion $\psi$ of the Bagger-Lambert theory belong to the ${\textbf{8}}_s$ and ${\textbf{8}}_c$ representations of the SO(8) R-symmetry and are 32-component spinors satisfying
\begin{equation}
\g^{\mu \nu \rho} \e = + \e^{\mu \nu \rho} \e , \quad \quad \g^{\mu \nu \rho} \psi = - \e^{\mu \nu \rho} \psi. \label{chirality}
\end{equation}
We assume that $\varepsilon_{012} = - \varepsilon^{012}$ and thus
\begin{align}
\e_{\mu \nu \lambda} \e^{\mu \rho \sigma} &= - (\delta^\rho_\nu \delta^\sigma_\lambda - \delta^\rho_\lambda \delta^\sigma_\nu ). \\
\e_{\mu \nu \lambda} \e^{\mu \nu \sigma} &= -2 \delta^\sigma_\lambda.
\end{align}
The following relations on $\mathcal{M}$ follow from the chirality constraint \eqref{chirality}
\begin{align}
\e^{\mu \nu \rho} \g_{\nu \rho} \e &= - 2 \g^\mu \e. \\
\e^{\mu \nu \rho} \g_\rho \e &= \g^{\mu \nu} \e .\\
\e^{\mu \nu \rho} \g_{\nu \rho} \psi &= + 2 \g^\mu \psi. \\
\e^{\mu \nu \rho} \g_\rho \psi &= - \g^{\mu \nu} \psi.
\end{align}
There is no analogue of the relations \eqref{chirality} on $\mathcal{N}$. However we do find the following information useful 
\begin{align}
\e_{\dot{1} \dot{2} \dot{3}} &= \e^{\dot{1} \dot{2} \dot{3}}. \\
\e_{\mud \nud \lambdad} \e_{\mud \rhod \sigmad} &= \delta_{\nud \rhod} \delta_{\lambdad \sigmad} - \delta_{\nud \sigmad} \delta_{\lambdad \rhod}. \\
\e_{\mud \nud \lambdad} \e_{\mud \nud \sigmad} &= 2 \delta_{\lambdad \sigmad}.\\
\g_{\mud \nud \rhod} &= \g_{\dot{1} \dot{2} \dot{3}} \e_{\mud \nud \rhod}. \\ 
(\g_{\dot{1} \dot{2} \dot{3}})^2 &= -1.
\end{align}
\newpage
\end{appendix}

\bibliographystyle{abbrv}
\bibliography{main}

\end{document}